\documentclass[letterpaper,12pt]{article}

\usepackage[margin=1in]{geometry}
\usepackage{authblk}
\usepackage[utf8]{inputenc} 
\usepackage[T1]{fontenc}    
\usepackage{hyperref}       
\usepackage{url}            
\usepackage{booktabs}       
\usepackage{amsfonts}       
\usepackage{nicefrac}       
\usepackage{microtype}      
\usepackage{graphicx}
\usepackage{natbib}
\usepackage{doi}

\usepackage{bbm}
\usepackage{multicol}
\usepackage{placeins}
\usepackage{subcaption}
\usepackage{amsmath}
\usepackage{amstext}
\usepackage{amssymb}
\DeclareMathOperator*{\argmin}{arg\,min}

\usepackage{setspace}
\title{A Kriging Metamodel with Adaptive Sampling for Seismic Evaluation of Podium Buildings}

\author[1]{Yuying Huang}
\author[2]{Zhiyong Chen}
\author[3]{Samuel W. K. Wong}
\affil[1,3]{Dept. of Statistics and Actuarial Science, Univ. of Waterloo}
\affil[2]{FPInnovations}
\date{}

\begin{document}
\maketitle

\begin{abstract}
In this paper, nonlinear time-history dynamic analyses of selected earthquake ground motions are conducted on designated wood-frame podium buildings and the resulting inter-story drifts are analyzed. We aim to construct a reliable region where performance-based seismic design criteria are met, such that a two-step analysis procedure can be used with high confidence. We develop a kriging metamodel with tailored adaptive sampling methods to achieve this goal in a computationally efficient manner. The input variables we consider are the normalized stiffness ratio and the normalized mass ratio of the podium building. We took a six-story wood frame built upon a one-story concrete podium as a case study for our methodology, where our results indicate that the two-step analysis procedure may be used with high confidence if its normalized stiffness ratio is at least 38 and its normalized mass ratio is between 0.5 and 1.5.
\end{abstract}

\section{INTRODUCTION\label{Intro}}

The wood-frame podium building is a common type of mid-rise timber structure made up of a wood-frame structure (often three to six stories) built upon a one- or two-story concrete podium. Due to inherent stiffness and mass irregularities changes along the building height, podium buildings require complicated dynamic analyses for seismic design. The 2015 edition of the National Building Code of Canada (NBCC) and the 2016 edition of the American Society of Civil Engineering Standard (ASCE 7) allow engineers to design podium buildings with a simple two-step analysis procedure when specific criteria are met. For podium buildings that meet the criteria, the two-step analysis procedure treats the upper wood-frame structure as an independent building on a fixed base, and the lower concrete structure as an independent building with forces from the upper structure at its top. Both the upper and lower structures can then be analyzed with the equivalent static force procedure. However, the criteria for utilizing the two-step analysis procedure in the 2015 NBCC commentary differ from those in ASCE 7. Further study is required to determine their appropriateness and efficacy in satisfying the intended seismic performance. 

To examine the criteria, \cite{chen2020criterion} performed nonlinear time-history dynamic analyses with selected earthquake ground motions following guidelines proposed by \cite{tremblay2015selection}. They investigated the relationship between the normalized mass and stiffness ratios (input variables) and the maximum inter-story drift (output variable). The normalized mass ratio (denoted by $[M]$) and normalized stiffness ratio (denoted by $[K]$) depend on the podium building's structure, and the maximum drifts determine whether the two-step analysis procedure can be used. Performance-based seismic design requires that the maximum inter-story drift is no greater than 4\%, with a non-exceedance probability of 80\% under maximum considered earthquake (MCE); see \cite{pang2010simplified} and \cite{van2010experimental}. The authors tested 15 earthquakes calibrated for Vancouver, British Columbia on 24 normalized mass-stiffness combinations (i.e., $[M]$, $[K]$ pairs). As a general criterion, they proposed that all podium building designs (i.e., podium buildings with different combinations of wood and concrete stories) should satisfy the performance-based seismic requirement when $[K]\geq 10[M]$. Although this criterion correctly classifies seismic reliability for the 24 inputs considered, further study is needed concerning its generalizability to the whole input space (i.e., all mass-stiffness combinations) and all podium building designs. 

Simulation is a widely-used tool to study complex underlying systems, e.g., for seismic response analysis in \cite{timber2022chen}. Evaluating the simulation function in real-life applications may be either computationally intensive or time-consuming. In this study, one nonlinear time-history analysis of 15 earthquakes for a given podium building takes approximately 30 hours. Practitioners often cannot afford to explore massive combinations or conduct a lengthy search in the input space. Metamodels can come to aid as fast surrogate models of the simulation function. They are fitted to the Input/Output data generated by the experiment with the simulation function and produce estimates for inputs that have not yet been simulated; see \cite{pellegrino2010fea} and \cite{jones2008finite}. A variety of metamodels such as low-order polynomial regression, kriging, and machine learning methods (e.g., artificial neural networks, support vector machines, etc.) have been proposed and extensively reviewed; see \cite{kleijnen2009Kriging} and \cite{asher2015review}.

It is desirable for the metamodeling process to generate an accurate surrogate model while exploiting as few samples as possible. Adaptive sampling techniques, which update the epistemic metamodel by adding new inputs repeatedly until a stopping rule is satisfied, are typically used in the application of metamodels. Various adaptive sampling schemes incorporating kriging have emerged as widely applied metamodeling techniques for expensive computational experiments. The reader may refer to \cite{fuhg2021state} for an insightful review of different adaptive sampling schemes.

This study develops a kriging metamodel with adaptive sampling to approximate the output inter-story drift as a function of the input normalized mass-stiffness combination.
The fitted metamodel gives predictions of the inter-story drift over the whole input space, and our immediate goal is to construct a reliable region where the criterion for a two-step analysis procedure is met with high confidence. To achieve this goal in a computationally efficient manner, we develop a tailored adaptive sampling method. As a specific case study, we focus on the building design that consists of a six-story wood-frame structure on a one-story concrete podium (hereafter $6+1$ podium building). More generally, our methodology could be adopted by researchers seeking to develop informative decision boundaries in other applications.

\section{METHODOLOGY\label{Method}}
In this section, we present our proposed algorithm and associated statistical methods used in this paper. Specifically, Section \ref{OK} introduces ordinary kriging, Section \ref{adaptive} formulates our customized adaptive sampling scheme and Section \ref{pipeline} concludes the algorithm pipeline. The algorithm we propose belongs to the class of single-selection adaptive sampling schemes that sequentially add one combination per iteration.

\subsection{Ordinary kriging metamodel \label{OK}}

 This paper focuses on kriging metamodels. Kriging originated from the work of South African mining engineer \cite{krige1951statistical} for use in geostatistics, and was later extended by \cite{matheron1963principles} and \cite{cressie1993statistics}. Several variants of kriging technique have been used in metamodeling, such as simple kriging, ordinary kriging and universal kriging, which differ in their assumptions and resulting modeling complexity. Among these, ordinary kriging (OK) is a popular and successful metamodel for simulations, and often achieves better accuracy compared to simple or universal kriging; see \cite{kleijnen2009Kriging}, and \cite{kleijnen2017regression}.  

Ordinary kriging is a spatial interpolation method that generates predictions as a weighted average of values at known combinations, i.e., 
\begin{equation}
    \hat{Y}(x_0)=\sum_{i=1}^n w_iY(x_i)
\end{equation}
where $Y(x_i), i=1,\dots, n$ are actual observations of measured combinations $\mathcal{S}=\{x_1,\dots, x_n\}$, $x_0$ is a new combination from the input space $\mathcal{D}$ and $\hat{Y}(x_0)$ is its kriging estimator. The OK estimator is an optimal linear estimator, in the sense that it is the unbiased estimator with the minimum mean-squared prediction error (MSPE). The weights $\{w_i\}$ are derived based on these unbiasedness and minimum MSPE properties.

For unbiasedness, the expected value of the OK estimator at $x_0$ must coincide with the unknown mean $\mu$ of the random field. This implies $E[\hat{Y}(x_0)]=E[\sum_{i=1}^n Y(x_i)]=\mu$; namely, $\sum_{i=1}^nw_i=1$. The OK estimator minimizes the MSPE, denoted by $\sigma_{x_0}^2$ (also known as the kriging variance),
\begin{align}
    \sigma_{x_0}^2&\equiv E[\hat{Y}(x_0)-Y(x_0)]^2 \nonumber\\
    &=-\sum_{i=1}^n\sum_{j=1}^nw_iw_j\gamma_{ij}+2\sum_{i=1}^nw_i\gamma_{i0},
\end{align}
where $\gamma_{ij}=\gamma \big(Y(x_i),Y(x_j) \big)=1/2Var[Y(x_i)-Y(x_j)]$, for $i\neq j$. Under the usual assumptions of second order stationarity and isotropy, the value of $\gamma_{ij}$
depends only on the distance $h$ between two combinations regardless of their location; $\gamma(h)$ is called the semi-variogram.

Given the unbiasedness condition, the kriging variance can be minimized by employing the method of Lagrange multipliers. Then the kriging equation in matrix notation becomes:
\begin{equation*}
\underset{\begin{bmatrix}
 0&\gamma_{12}&\dots&\gamma_{1n}&1\\
\gamma_{21}&0&\dots&\gamma_{2n}&1\\
\vdots&\vdots&\ddots&\vdots&\vdots\\
\gamma_{n1}&\gamma_{n2}&\dots&0&1\\
1&1&\dots&1&0
\end{bmatrix} \cdot}{\Gamma}
  \underset{\begin{bmatrix}
w_1\\w_2\\\vdots\\w_n\\-\lambda
\end{bmatrix}=}{W}
\underset{\begin{bmatrix}
\gamma_{10}\\\gamma_{20}\\\vdots\\\gamma_{n0}\\1\end{bmatrix}}{D}
\end{equation*}
where $\lambda$ is the Lagrange multiplier.
We can express the estimated weights and estimated kriging variance in matrix forms as follows:
\begin{equation}\label{variogram}
W=\Gamma^{-1}D,~~~ 
    \sigma_{x_0}^2=D^T\Gamma^{-1}D
\end{equation}
As suggested by Eq.(\ref{variogram}), a semi-variogram is required for computing $W$ and $\sigma_{x_0}^2$.

In applications, the semi-variogram is generally unknown and it is necessary to select a theoretical semi-variogram model to fit the empirical values. Some commonly used variogram model families are considered in this paper, see, e.g., \cite{webster2007environmental, li2022spatial}. They are the Bounded linear model, the Spherical model, the Exponential model, and the Gaussian model, which are parametrized as follows.
\begin{itemize}
\item[(i)] Bounded linear model: \begin{equation*}
    \gamma^{lin}_{(a,b)}(|h|)=\begin{cases}C_0+b(\dfrac{|h|}{a}),~0<|h|\leq a \\
C_0+b,~\text{otherwise} \end{cases} 
    \end{equation*}
\item[(ii)] Spherical model: \begin{equation*}
    \gamma_{(a,b)}^{sph}(|h|)=\begin{cases}C_0+b\big( \dfrac{3|h|}{2a}-\dfrac{|h|^3}{2a^3}\big),~0< |h|\leq a \\
C_0+b,~\text{otherwise} \end{cases} 
    \end{equation*}    
\item[(iii)] Exponential model: \begin{equation*}
    \gamma_{(a,b)}^{exp}(|h|)=C_0+b\big( 1-exp\big ( -\dfrac{|h|}{a}\big )\big ),~|h|>0
    \end{equation*} 
\item[(iv)] Gaussian model: \begin{equation*}
\gamma_{(a,b)}^{gau}(|h|)=C_0+b\big( 1-exp\big ( -\dfrac{|h|^2}{a^2}\big )\big ),~|h|>0
\end{equation*}
\end{itemize}
Here, $\gamma_{(a,b)}(|h|)$ denotes the semi-variogram for each model as a function of distance $|h|>0$ between two combinations, with $C_0, a, b \geq 0$ as the parameters: $C_0$ represents the nugget effect, $a$ represents the range parameter, and $b$ represents the partial sill value. For a chosen variogram function with fitted parameters, the estimated kriging variance of the combination $x_0$ is denoted by $\hat{\sigma}^2_{x_0}$.

\subsection{Proposed adaptive sampling method\label{adaptive}}

Our goal is to determine the acceptable regions of combinations where their inter-story drifts meet the performance-based seismic design criterion, for using a two-step analysis procedure.
By sequentially adding new measured combinations, we seek to decrease overall uncertainty while focusing our adaptive sampling effort on subregions whose drifts are likely to be around the 4\% threshold. The selection of the next combination will depend on the refinement criterion (denoted by RC), which is employed as a ``loss function''. In single-selection schemes, the next combination to be measured is the one that optimizes the RC. A critical question is how to design the RC to achieve our goal.

The first step is to characterize the global kriging uncertainty in terms of the pointwise (i.e., at individual combinations) kriging variance.
One commonly used metric is the average kriging variance (AKV), defined as
\begin{equation*}
A(\mathcal{S})=\int_{\mathcal{D}} \sigma^2_{x}w(x) dx \approx \sum_{\mathcal{D}} \sigma^2_{x}w(x)
\end{equation*}
where $w(x)$ is the weight assigned to combination $x$, $\sigma^2_{x}$ is the true kriging variance at combination $x$ conditional on observing $\mathcal{S}$. As suggested in \cite{zhu2006spatial}, the AKV integral is approximated by summation; moreover, as the combination's true variance is usually unknown, we may substitute $\sigma^2_{x}$ with its estimate $\hat{\sigma}^2_{x}$ (via the fitted variogram).

The second step is to formulate an RC in terms of the AKV over the unobserved combinations $x\in \mathcal{D}/\mathcal{S}$. The next combination to be measured, $x_{n+1}$, can then be chosen according to
\begin{equation*}
x_{n+1}=\underset{x\in \mathcal{D}/\mathcal{S}}{\argmin} \sum_{x_0\in \mathcal{D}/\{\mathcal{S},~ x\} } \hat{\sigma}^2_{x_{0}} w(x_0)
\end{equation*}
Note that $x_{n+1}\in \mathcal{D}/\mathcal{S}$ will be chosen prior to $Y(x_{n+1})$ being measured. To do this, we use the fitted variogram based on $\{Y(x_1), \ldots, Y(x_n)\}$ throughout the computation, keeping it fixed when each new combination $x\in \mathcal{D}/\mathcal{S}$ is hypothetically included.

The third step is to select an appropriate weight function $w(x)$. For example, if we take $w(x) = 1$, the RC weights every combination equally and hence only reduces global uncertainty.  With limited computing resources, this may not be most efficient for determining the reliable region. Rather, we develop a tailored $w(x)$ that enables us to weight the combinations according to their importance. 

If the metamodel gives us high confidence that a combination is either well above or well below the threshold value of $4\%$, we do not need to prioritize it for measurement. Specifically, we calculate the confidence intervals (CI) of kriging estimates to quantify their uncertainty. We then attach higher importance to the combinations that have CIs containing $4\%$. This leads to the weight function

$$w(x) = \mathbbm{1}(\hat{Y}(x))=\begin{cases}
0,& \text{if}~CI_{\alpha, lower}(x)>d,\\
&\text{or}~CI_{\alpha, upper}(x)\leq d\\
1,&\text{otherwise}
\end{cases}$$
where $CI_{\alpha, lower}$ and $CI_{\alpha, upper}$ are the lower and upper bounds of the kriging CI for combination $x$ under the confidence level $(1-\alpha)\times100\%$,
and $d$ is the threshold value. In our paper, we take $\alpha=0.1$ and $d=4\%$, and the $90\%$ CI for $Y(x_0)$ is constructed using a Normal approximation: 
$$\big[\hat{Y}(x_0)-1.645\hat{\sigma}_{x_0},\hat{Y}(x_0)+1.645\hat{\sigma}_{x_0} \big].$$ 
Finally, our proposed RC is established as follows:
\begin{equation}\label{RC}
x_{n+1}=\underset{x\in \mathcal{D}/\mathcal{S}}{\argmin} \sum_{x_0\in \mathcal{D}/\{\mathcal{S},~ x\} } \mathbbm{1}\big(\hat{Y}(x_0) \big)\hat{\sigma}^2_{x_{0}}.    
\end{equation}

Our method relies on the idea that large uncertainties in metamodel approximation occur in regions where the predicted variances are large. When few combinations have been measured, our RC tends to prioritize reducing global uncertainty to obtain a comprehensive understanding of the input space; as more combinations are measured and CIs become more informative, 
the sampling
focuses on sub-regions whose CIs contain the threshold value, i.e., where more measurements are needed to determine the reliable region. Kriging variances and CIs, as direct byproducts of the kriging metamodel, are utilized in this RC. There are different adaptive sampling techniques based on kriging variance in \cite{jones1998efficient, lam2008sequential, sobester2005design}. These adaptive techniques are commonly used in optimization problems, where the goal is to locate the global minimum value over the whole input space. In contrast,
we aim to accurately classify combinations into regions above or below a threshold value. To the best of our knowledge, this research goal has rarely been discussed and explored in the field of kriging metamodels incorporating adaptive techniques, which motivated us to develop and propose our own adaptive method.

\subsection{Algorithm pipeline\label{pipeline}}
Based on the RC above, the sequential algorithm for the development of our metamodel is presented as follows:
\begin{itemize}
    \item[1.] To initialize the adaptive sampling process, a set of initial measurements at the combinations $\mathcal{S}^{n}=\{x_1,\dots,x_n\}$ are needed.  These could be obtained from prior study, randomly chosen, or follow a space-filling design; see \cite{fuhg2021state}. Additionally, the ranges of the input space need to be determined.
    \item[2.] Fit the metamodel with the current dataset of measurements at $\mathcal{S}^n$. Four theoretical variogram models (see Section \ref{OK}) are fitted to the data and the one with the lowest mean-squared error is chosen.
\item[3.] Find the next combination $x_{n+1}$ from $\mathcal{D}/\mathcal{S}^{n}$ that optimizes the customized RC as in Eq.(\ref{RC}).
\item[4.] Update the dataset of measured combinations, $\mathcal{S}^{n+1}=\{\mathcal{S}^n, x_{n+1}\}$.
\item[5.] Repeat Steps 2-4 until a stopping rule is reached. 
\end{itemize}

As we apply the RC in $\mathcal{D}/\mathcal{S}$, our iterative algorithm will naturally come to stop when all combinations in the input space are either measured or have predicted confidence intervals that do not contain 4$\%$.
The number of iterations the algorithm takes to naturally stop (denoted as $n_{stop}$) varies and is usually unknown, so we also set a maximum number of iterations as 50 to meet computational time constraints. Thus, our stopping rule concludes adaptive sampling after $\min\{n_{stop}, 50\}$ iterations.

\section{APPLICATION\label{App}}

This section presents the application of our proposed method through a case study.
We introduce the experimental setup in Section \ref{design}, then discuss the results and visualize the final metamodel predictions in Section \ref{result}.

\subsection{Experimental setup\label{design}}
As a case study, we focus on the $6+1$ podium building design. Each input combination is a $[M],[K]$ pair, defined following Chen and Ni's framework:
$$
[M]=\dfrac{M_Cn_C}{M_Wn_W},~~[K]=\dfrac{K_C/n_C}{K_W/n_W},
$$
where $M_W$ and $K_W$ denote the mass and stiffness respectively, of the bottom story of the upper wood-frame; $M_C$ and $K_C$ denote the mass and stiffness respectively, of the top story of the lower concrete structure. Here, $n_C$ and $n_W$ denote the number of stories in the concrete and wood structures, respectively, which are fixed as $n_C=1$ and $n_W=6$ for our building type. We also hold $M_W=2389$ (kg) and $K_W=4.4$ (kN/mm) fixed, so that different values of $[M]$ and $[K]$ are obtained by adjusting $M_C$ and $K_C$.

During earthquake simulation, inter-story drifts between each adjacent story are produced; for a $6+1$ podium building, there will be seven inter-story drifts (including the drift at the ground level). We analyze the 15 selected earthquakes as in \cite{chen2020criterion} and calculate the maximum inter-story drift per earthquake. We then take the fourth-largest of these 15 individual maximum drifts as our response variable: if this value does not exceed 4\%, then at least 12 out of 15 of the earthquakes (80\%) have maximum drifts within 4\%, which satisfies the seismic criterion.

The region of interest, $\mathcal{D}$, is spanned by the normalized mass and stiffness ratios. For our study, $[M]$ ranges from 0.5 to 6 with a stride of 0.5, and $[K]$ ranges from 1 to 60 with a stride of 1, which yields a total of 720 measurable combinations covering the area $[0.5,6]\times[1,60]$. \cite{chen2020criterion} initially considered ranges for variables $[K]$ and $[M]$ were both $(0,60]$; however, their results indicated that two-step analysis procedure will always fail when $[K] < 10[M]$. Thus, we may exclude the region where $[M] \geq 6$ and $[K] \leq 60$ from consideration.

We initialize our adaptive sampling process with the same combinations as those simulated in Chen and Ni's experiment. These 12 combinations of $[M],[K]$ are plotted as open circles in Figure \ref{fig:first}; note that while their experiment included 24 measured combinations, only those that lie within our region of interest are relevant. To help ensure that the kriging isotropic assumption is reasonable, we examined plots of the directional fitted variograms based on these 12 initial combinations.  We found that isotropy is reasonably met  by rescaling $[K]$ (y-axis) to $[K]/10$, and so kriging is performed under this rescaling. The final results will be shown on the original scale.

\FloatBarrier
\subsection{Results\label{result}}

\begin{figure}[!htbp]
\centering
\begin{subfigure}{\linewidth}
\centering
    \includegraphics[scale=0.4]{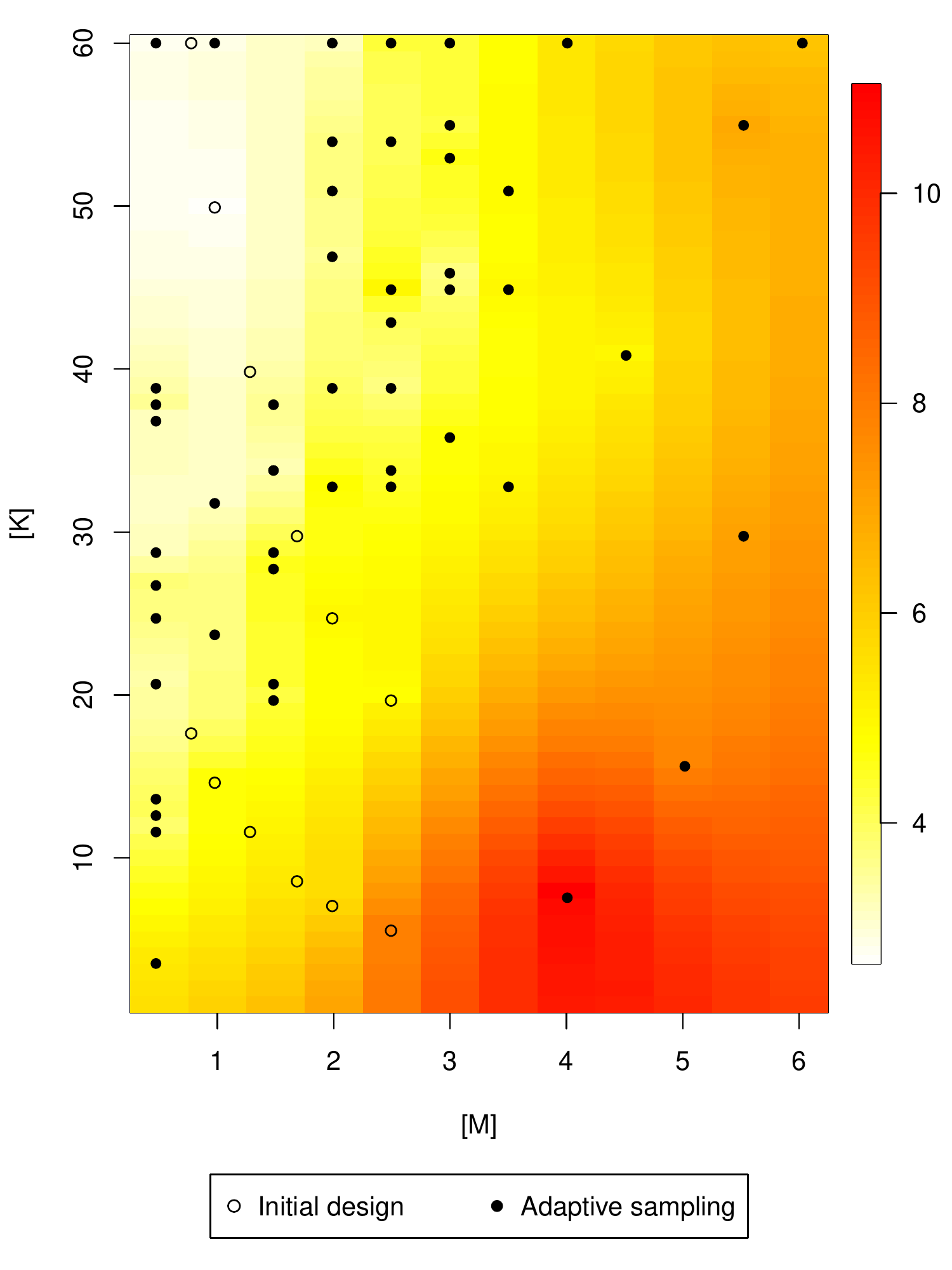}
    \caption{Kriging predictions and sampled combinations.}
    \label{fig:first}
\end{subfigure}
\hfill
\begin{subfigure}{\linewidth}
\centering
    \includegraphics[scale=0.4]{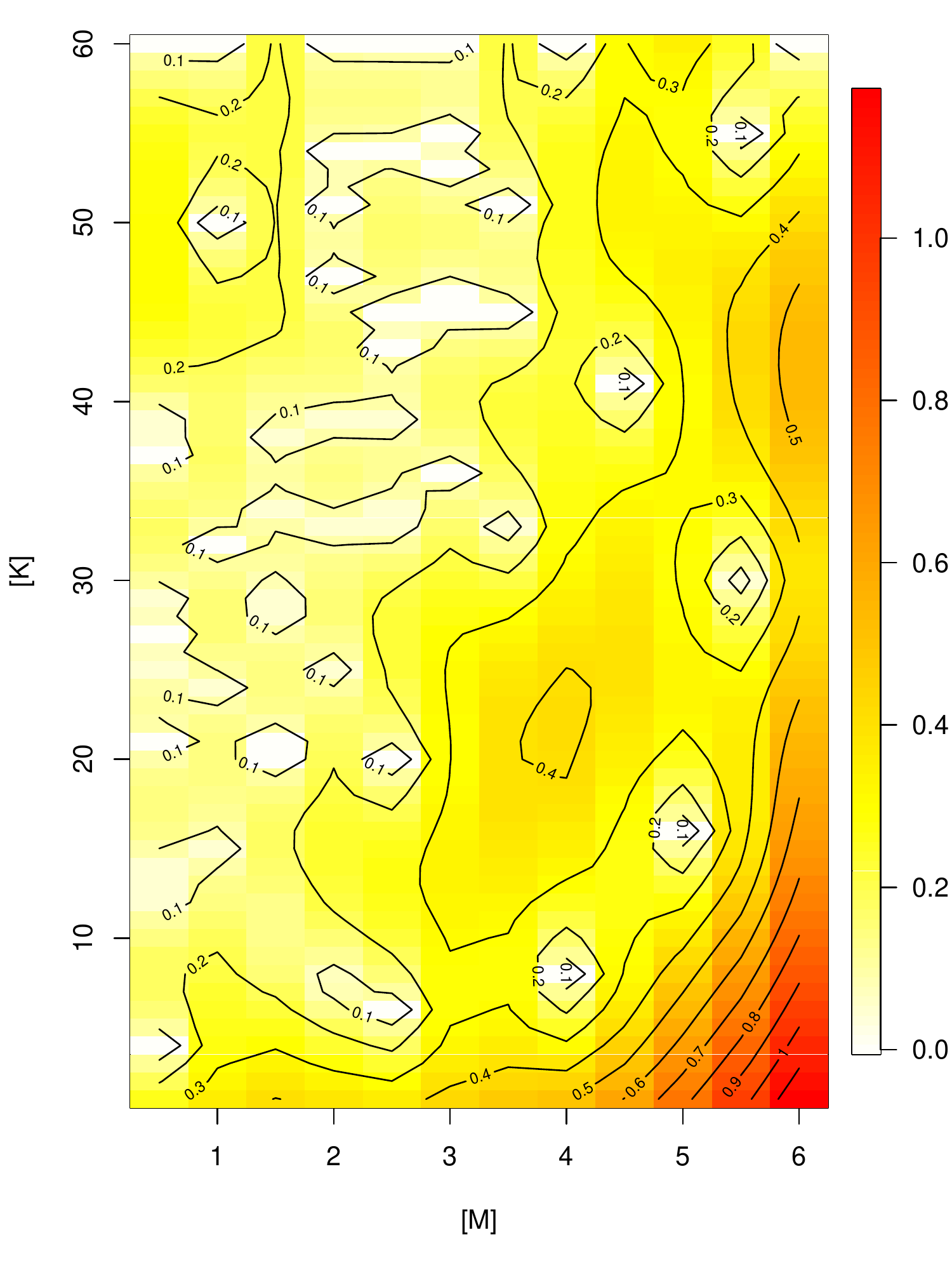}
    \caption{Kriging predictor variance.}
    \label{fig:second}
\end{subfigure}
\label{fig:figures}
\caption{Kriging metamodel predictions based on the final total of 62 measurements obtained by adaptive sampling (top panel) and corresponding kriging variance (bottom panel).}
\end{figure}

Our sequential algorithm reached 50 iterations, the maximum allowed under our stopping rule. As seen in Figure \ref{fig:first} via the solid dots, more adaptive combinations are sampled in the upper-left region, while the bottom-right region is sparsely sampled. The algorithm searches most intensively where the kriging metamodel predictions (shown via the shaded colors) tend to be close to the threshold drift of 4\% (yellow) while exploring little in regions where drifts are well-above 4\% (red) or well-below 4\% (white). The kriging predictor variances over the entire input space are illustrated in Figure \ref{fig:second} with shaded colors and contour lines superimposed.
Predictor variances tend to be larger in regions with fewer measurements (e.g., bottom-right), and smaller if a nearby combination has been measured. 
 
\begin{figure}[h]
  \centering
  \includegraphics[scale=0.4]{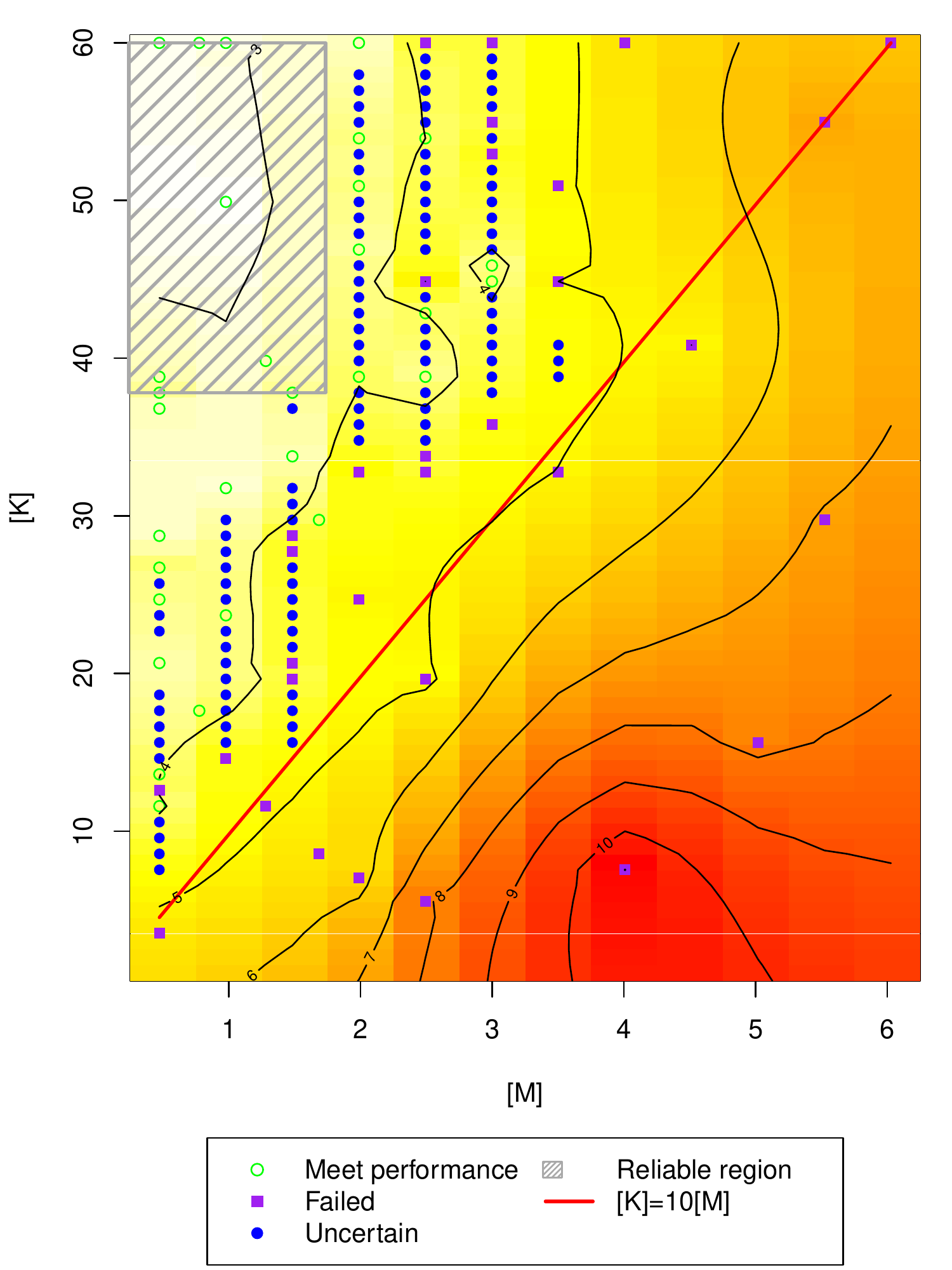}
\caption{
Kriging predictions with contour lines and labelled combinations. The empirical criterion proposed by \cite{chen2020criterion} is plotted together with our reliable region for the $6+1$ podium building. \label{fig:third}}
\end{figure}

Figure \ref{fig:third} provides a more detailed look at the measurements and final kriging metamodel. Green circles indicate measurements no greater than 4\%, and purple squares indicate measurements exceeding 4\%. The locations of these green and purple markers suggest that the performance-based seismic design criterion is met with a lower normalized mass ratio and a higher normalized stiffness ratio. Furthermore, blue dots indicate unmeasured combinations where predictions have 90\% CIs with upper limits that exceed 4\%. Ideally,  combinations sampled closely to each other will have similar outputs. However, in practice there was higher than expected variability in the outputs of earthquake simulations; mathematically, this leads to a fitted variogram with a nugget effect of 0.025, which implies that neighbouring combinations are not perfectly correlated. Consequently, the metamodel has prediction uncertainty at all unmeasured combinations. In particular, some blue dots are directly adjacent to green measurements; these combinations may have metamodel predictions $\le 4\%$ but their confidence is not high.

As a practical recommendation from this case study, we may construct a conservative reliable region in which predictions are $\le 4\%$ with high confidence. Two conditions should be satisfied by combinations in the reliable region: (i) metamodel predictions $\le 4\%$, as visually indicated by the region left of the 4\% contour line; (ii) 90\% CIs with upper limits that do not exceed $4\%$, to account for the uncertainty of kriging predictions (i.e., ``uncertain'' combinations are excluded). The largest such contiguous reliable region is shaded with dark grey lines in Figure \ref{fig:third}, in which $[K]$ is no less than 38 and simultaneously $[M]$ is no less than 0.5 but no greater than 1.5. With more computational resources and additional simulation, our reliable region might be expanded by obtaining more measurements at the ``uncertain'' combinations.

In contrast to the original $[K]/[M]\geq 10$ criterion recommended in \cite{chen2020criterion}, our experiments suggest that the $6+1$ podium building may require a more stringent reliable region. The red solid line in Figure \ref{fig:third} shows Chen and Ni's $[K]=10[M]$ boundary; however, simulations from this case study indicate that the $[K]/[M]\geq 10$ region (upper-left of the red line) contains both ``meet performance'' and ``failed'' combinations. Furthermore, different podium building designs (e.g., $4+1$, $5+1$, etc.) will need to be studied to determine whether individually-tuned criteria (reliable regions) are needed. Finally, depending on study objectives, e.g., in structural engineering applications, a lower 75\% confidence level might be preferred over the conservative 90\% CI used here, which would result in fewer combinations marked as ``uncertain'' and a larger reliable region.

\FloatBarrier
\section{CONCLUSIONS\label{Conclusion}}
This paper used earthquake simulations to explore the seismic responses of podium buildings, to derive a reliable region for applying the two-step analysis procedure. In our case study, the performance-based seismic design criterion was satisfied when the maximum inter-story drift of the podium building did not exceed 4\% among 12 out of 15 selected earthquakes. Our goal was to identify combinations in the normalized mass-stiffness input space that met this criterion. We employed the kriging geostatistics technique, in conjunction with a tailored adaptive sampling scheme, to increase the effectiveness of the metamodeling process. For the $6+1$ podium building, a conservative reliable region based on the measurements from this study is $38\leq [K]\leq 60$ and $0.5\leq[M]\leq 1.5$, which is more stringent than \cite{chen2020criterion} but could be refined by further simulation. The practical utility of our proposed statistical method was demonstrated, as the adaptive sampling algorithm focused on reducing prediction uncertainty at combinations near the 4\% threshold value. We found that the inherent measurement errors of earthquake simulation were larger than expected, which potentially affected the evolution of the sampling scheme and prediction uncertainty of the metamodel. We are motivated to address this shortcoming in the kriging metamodel and generalize our methods to other podium building designs in our future work. 

\FloatBarrier
\section{REFERENCES}
\bibliographystyle{unsrtnat}
\bibliography{bib}

\end{document}